\documentclass[a4paper]{article}

\usepackage[english]{babel}
\usepackage[utf8x]{inputenc}
\usepackage[T1]{fontenc}

\usepackage[a4paper,top=3cm,bottom=2cm,left=3cm,right=3cm,marginparwidth=1.75cm]{geometry}

\usepackage{amsmath}
\usepackage{amssymb}
\usepackage{graphicx}
\usepackage{url}
\usepackage[colorinlistoftodos]{todonotes}
\usepackage[colorlinks=true, allcolors=blue]{hyperref}
\usepackage{cite}

\numberwithin{equation}{subsection}

\title{Canonical analysis of Kalb-Ramond-Proca duality}

\author{Felipe A. da Silva Barbosa}

\begin{document}

\date{}

\maketitle

\begin{center}
    Departamento de Física,
Universidade Estadual Paulista, UNESP,
 Guaratinguetá, SP, Brazil
\end{center}

\begin{center}
    felipe-augusto.barbosa@unesp.br
\end{center}

\begin{abstract}
    It is shown that the canonical quantization of the free massive Kalb-Ramond and Curtright-Freund Lagrangians leads to the same theory obtained from the canonical quantization of the free Proca and Klein-Gordon Lagrangians. The duality in the presence of interaction is explored in the context of the Feynman rules and beyond. It is pointed out that the equivalence between massive dual models without gauge symmetry is rooted in an ambiguity of coordinate choices.
\end{abstract}

\section{Introduction}

     In the past few years there has been a renewed interest in self-interacting massive spin-1 theories in the context of modified gravity ~\cite{Tasinato:2014eka,Heisenberg:2014rta,deRham:2020yet,deRham:2021efp}. These self-interacting theories were built up with the Proca Lagrangian \cite{Proca:1936fbw} as the starting point. There are several claims in the literature \cite{Kawai:1980qq,Quevedo:1996uu,Quevedo:1997jb,Smailagic:2001ch,Kuzenko:2020zad,Hjelmeland:1997eg,DeGracia:2017spm} stating that the Kalb-Ramond \cite{PhysRevD.9.2273} and Proca Lagrangians are dual to each other, that is both theories carry the same number of degrees of freedom and describe the same physics. We will refer to the equivalence between Proca and an antisymmetric tensor as Kalb-Ramond-Proca duality, but it should be noticed that Kalb and Ramond were not the first to realize this equivalence \cite{Ivanov:2016lha,Ogievetsky:1966eiu}. Recently, a new massive model dual to Proca based on a symmetric tensor has also been found \cite{Dalmazi:2011df}. Naturally, one can ask what theories can be constructed with these dual models, explored for Kalb-Ramond in \cite{Heisenberg:2019akx}, and how they are related to Proca Lagrangians. We also have dual massive models for other spins like the Curtright-Freund vector model for spin-0 and tensor model for spin-2 \cite{Curtright:1980yj,Casini:2002jm,Gonzalez2008DualityFM}. It has been suggested that such alternative models of massive spin-2 could provide new theories of massive gravity \cite{Alshal:2019hpk}. One could also ask what relation theories constructed with these models have with the usual Klein-Gordon and Fierz-Pauli Lagrangians. 
     
     Starting with a massive dual model like Kalb-Ramond,  a dual model without gauge symmetry, the usual approach to defend the duality between models is to use the parent action method: we propose an action that depends on a vector and an antisymmetric field, in the case of Kalb-Ramond-Proca duality, one can integrate out the vector field or the rank-2 tensor to obtain the Kalb-Ramond or Proca Lagrangians, respectively. Since both Lagrangians came from the same parent action it is argued that they describe the same physics. It is possible to construct parent actions for other dual massive models so that one can use the same arguments to defend the equivalence. Although the investigation of equivalence between dual models started in flat space there are also extensions to curved backgrounds as well \cite{article,PhysRevD.78.084024}. It was noted that some of the dualities in string theory and supersymmetric gauge theories could be understood as canonical transformations \cite{Lozano:1996sc}. For Kalb-Ramond-Proca (KR-P) duality there are also some canonical formulations, but focusing on the canonical structure of Kalb-Ramond in higher dimensions \cite{Escalante:2014mva} and the duality between gauge theories of spin-1 \cite{Harikumar:1999tm}. We feel that the content of duality relations between massive models without gauge symmetry has not been investigated in the language of the canonical formalism. We expect that this paper can fill the gap. Moreover, using cosmological perturbation theory it was suggested that Kalb-Ramond and Proca may not be equivalent \cite{Hell:2021wzm}, we also hope that this work can clarify a point that was overlooked. 
    
     The goal of this paper is to use the canonical formalism to propose a new perspective on the duality between massive models. This new perspective is justified by its generality, it gives a framework to understand dual massive models for any spin and different dual models for the same spin. We will be working with the Kalb-Ramond and the vector spin-0 model as examples, but the arguments are intended to be valid for other dual massive models without gauge symmetry as well. We start with the free theories. Straightforward canonical quantization of Kalb-Ramond and Curtright-Freund shows that they describe the same physical system of Proca and Klein-Gordon, that is the Hilbert space of massive spin-1 and spin-0 particles. In particular, the Kalb-Ramond and Curtright-Freund fields are proportional to the Field strength of Proca and the derivative of the Klein-Gordon field. The relation between free theories immediately motivates the definition of duality between Feynman rules: given a Kalb-Ramond or Curtright-Freund Lagrangian which generates a set of Feynman rules we expect that there is a Proca or Klein-Gordon Lagrangian such that the same set of Feynman rules can be derived. This argument is formulated in the context of interaction picture and path integral, where its generality is clearer. We propose an interaction and use the canonical formalism to show that the equivalence between Kalb-Ramond and Proca goes beyond perturbation theory. The key point is that the interacting Hamiltonians are still related by canonical transformations of the Heisenberg picture operators. We propose a new way of using the parent action method, based on coordinate changes. Motivated by the coordinate changes between the Hamiltonians of dual models, and by the fact that the use of parent actions in the functional generator is not straightforward in the presence of interactions. Finally, using the parent action we propose that the non-perturbative equivalence may be valid for more general interactions. 
   
     The notation used here follows \cite{weinberg_1995} and is assumed that the reader is familiar with the canonical quantization procedure of constrained systems formulated by Dirac \cite{weinberg_1995,dirac2001lectures,hanson1976constrained}.

 \section{Free-theory}
    
   The duality between the free spin-1 duality between Proca and Kalb-Ramond and the spin-0 duality between Curtright-Freund and Klein-Gordon are explored. 
    
    \subsection{Kalb-Ramond}
    
      First, we will work out the Kalb-Ramond theory \cite{PhysRevD.9.2273}:
\begin{align}\label{KR}
    S_{KR} = \int d^4 x \left[\frac{\left(\partial^\mu B_{\mu \nu} \right)^2 }{2} + \frac{m^2}{4} B_{\mu \nu} B^{\mu \nu} \right],
\end{align}
showing that it has massive spin-1 content. Our signature is $(-,+,+,+)$ and $B_{\mu \nu} = - B_{\nu \mu}$. In order to see the spin-1 content in the canonical formalism we derive the momenta 
\begin{align}
    \Pi^{\alpha \beta}  \equiv \frac{\partial \mathcal{L}_{KR}}{\partial \dot{B}_{\alpha \beta}  }  = \partial_\mu B^{\mu \alpha} \delta^\beta_0 - \partial_\mu B^{\mu \beta} \delta^\alpha_0,
    \end{align}
where the $[i,j]$ components are zero, $\Pi^{ij} = 0$, and thus constraints. The Hamiltonian is 
\begin{align}
    H_{KR} = \int d^3 x \left[ \frac{1}{2} \Pi^{0 i} \Pi^{0 i} - B_{ji} \partial^j \Pi^{0 i}
    + \frac{1}{2} \partial^j B_{0 j} \partial^k B_{0 k } - \frac{m^2}{4} B_{\mu \nu} B^{\mu \nu} + \lambda_{ij} \Pi^{i j} \right],
\end{align}
with $\lambda_{ij}$ some Lagrange multipliers. Since the canonical variables are antisymmetric tensors the Poisson brackets are 
\begin{align}
    \{ B_{\mu \nu}(\boldsymbol{x},t) \, , \, \Pi^{\alpha \beta}(\boldsymbol{y}, t) \} = \left( \delta^\alpha_\mu \delta^\beta_\nu - \delta^\alpha_\nu \delta^\beta_\mu \right) \delta^3 (\boldsymbol{x}- \boldsymbol{y} ) .
\end{align}
   The consistency of the constraint $\Pi^{ij} = 0$ generates the new constraint
\begin{align}\label{xi_ij}
    \xi^{ij} \equiv  \{ \Pi^{i j} \, ,  \, H \}  =  \partial^i \Pi^{0j} - \partial^j \Pi^{0 i } + m^2 B^{i j } = 0,
\end{align}
 and $\dot{\xi}^{ij}=0$ fixes $\lambda_{ij}$, ending the algorithm.
 
 The constraints  of the theory 
\begin{align}\label{relationsKR}
    \Pi^{i j } =  0, \qquad \xi^{ij} = 0 ,
\end{align}
are second class. With six second class constraints, we are left with $12 - 6 = 6$ degrees of freedom in the Hamiltonian formalism, or $3$ in the Lagrangian formalism, consistent with spin-1 content.  However, just counting degrees of freedom is not enough to convince us that Kalb-Ramond and Proca describe the same physical system. Fortunately, it is possible to go further.

The reduced Hamiltonian is 
\begin{align}\label{hkr}
H_{KR}^R = \int d^3 x \left[ \frac{1}{2} \Pi^{0i} \Pi^{0i} + \frac{1}{2} \partial^j B_{0 j} \partial^l B_{0 l} + \frac{m^2}{2} B_{0j} B_{0j} +\frac{m^2}{4} \left( \partial^i \Pi^{0j} - \partial^j \Pi^{0i}\right)^2 \right], 
\end{align}
and from the relations \eqref{relationsKR} we conclude that the Dirac-brackets of independent variables are
\begin{align}\label{diracbc}
   \{ B_{0 i} (\boldsymbol{x},t) \,  , \,  \Pi^{0 j } ( \boldsymbol{y} , t) \}_D = \delta^j_i \delta^3 ( \boldsymbol{x}- \boldsymbol{y} ) .
\end{align}
Recall that given a set of constraints $\chi_N = 0 $ the Dirac-bracket is
\begin{align}\label{dirac-brack}
    \{ A \, , \, B \}_D \equiv \{ A\, ,B\, \} - \{A \, ,\,  \chi_N \}  (C^{-1})^{NM} \{ \chi_M \, , \, B \},
\end{align}
where $(C^{-1})^{NM}$ is the inverse of the constraint matrix $C_{NM} \equiv \{ \chi_N \, , \, \chi_M \}$. 

The quantization procedure consists of promoting the reduced Hamiltonian to the Hamiltonian operator, the constraint relations to equalities among operators and the commutators among fields are given by the Dirac-brackets through 
\begin{align}\label{KR-comm}
    \left[ B_{0 i} (\boldsymbol{x},t) \,  , \,  \Pi^{0 j } ( \boldsymbol{y} , t) \right] = i \, \{ B_{0 i} (\boldsymbol{x},t) \,  , \,  \Pi^{0 j } ( \boldsymbol{y} , t) \}_D = i \,    \delta^j_i \delta^3 ( \boldsymbol{x}- \boldsymbol{y} ).
\end{align}
The Hamiltonian equations of motion, coming from the Heisenberg equations reads
\begin{align}\label{eqmH}
  \dot{B}_{0 i}   =  \Pi^{0 i } + \partial^j B_{j i} \, , \qquad \dot{\Pi}^{0 i} =  \partial^i \partial^j B_{0 j} - m^2 B_{0 i}
\end{align}
and with the constraints $\xi^{ij} = 0$ all together return the covariant equations of \eqref{KR}
\begin{align}\label{eqmKR}
    m^2 B_{\mu \nu} - \partial_\mu \partial^\alpha B_{\alpha \nu} + \partial_\nu \partial^\alpha B_{\alpha \mu} = 0.
\end{align}
 The divergence of \eqref{eqmKR} gives the wave equation
\begin{align}\label{eqmKR2}
    ( \Box - m^2 ) \partial^\alpha B_{\alpha \nu}  = 0 .
\end{align}
The most general solution for $B_{\mu \nu}$ that satisfies \eqref{eqmKR} is
\begin{align}\label{KR3}
    B_{\mu \nu} = \frac{F_{\mu \nu}(A)}{m}, \qquad F_{\mu \nu}(A) \equiv \partial_\mu A_\nu - \partial_\nu A_\mu ,
\end{align}
where $A_\nu $ is the most general solution for $\partial^\alpha B_{\alpha \nu}$ in \eqref{eqmKR2}:
\begin{align}\label{eqmdualKR}
 \partial^\alpha B_{\alpha \nu} \equiv m A_\nu, \qquad ( \square - m^2 )   A_\nu = 0, \qquad \partial^\nu A_\nu =0.
\end{align}
We can prove this by contradiction. If \eqref{KR3}  were not the most general solution for \eqref{eqmKR} then the general solution would assume the form
\begin{align}
    B_{\mu \nu} = \frac{F_{\mu \nu}}{m} + \tilde{B}_{\mu \nu},
\end{align}
where $\tilde{B}_{\mu \nu}$ is linearly independent of $F_{\mu \nu}$, that is
\begin{align}\label{LI-B-F/m}
    a \frac{F_{\mu \nu}}{m}  + b \tilde{B}_{\mu \nu} = 0 \quad \forall \, \, x \quad\Leftrightarrow \quad a=b=0 .
\end{align}
Taking the divergence of this equation and using \eqref{eqmdualKR} we have
\begin{align}
   a\,  m A_\nu  +  b\,  \partial^\mu \tilde{B}_{\mu \nu} = 0 \quad \forall \, \, x \quad \Leftrightarrow \quad a=b=0,
\end{align}
since $\tilde{B}_{\mu \nu}$ is a solution for \eqref{eqmKR} it also satisfies \eqref{eqmKR2}, so that $\partial^\alpha \tilde{B}_{\alpha \nu}$ is a new solution for \eqref{eqmKR2} linearly independent of $A_\nu$, thus contradicting our assumption that $A_\nu$ was the most general solution\footnote{We can obtain the same result using \eqref{eqmKR2} to show that $(\square -m^2)B_{\mu \nu}=0$, then we can expand for $B_{\mu \nu}$ in terms of its plane wave solutions and use \eqref{eqmKR} to define the Fourier coefficients.}. 

The general solution for a vector field satisfying \eqref{eqmdualKR} can be expressed in terms of a Fourier transform 
\begin{subequations}\label{defsA}
\begin{align}\label{generalA}
     A_\mu (x) = (2 \pi)^{-3/2} \sum_{\sigma} \int d^3 p (2 p^0)^{-1/2} \{ e^{\mu}(\boldsymbol{p}, \sigma ) a( \boldsymbol{p}, \sigma) &e^{i p \cdot x} \nonumber\\
     & + e^{* \mu} (\boldsymbol{p}, \sigma)  a^{\dagger} (\boldsymbol{p}, \sigma) e^{-i p \cdot x} \},
\end{align}
where
\begin{align} \label{defA1}
  & p^0 \equiv \sqrt{\boldsymbol{p}^2 + m^2 } ; \\ \label{defA2}
   & p_\mu e^{\mu}(\boldsymbol{p}, \sigma) = 0 , \qquad \sum_{\sigma} e^{\mu}(\boldsymbol{p}, \sigma) e^{\nu *}(\boldsymbol{p}, \sigma)  = \eta^{\mu \nu} + p^\mu p^\nu /m^2.
\end{align}
\end{subequations}
with $\sigma=1,0,-1$; and $a(\boldsymbol{p}, \sigma) $ are operator coefficients. In order to fix these operator coefficients we use the canonical commutation relations \eqref{KR-comm}: note that from \eqref{eqmH} and \eqref{KR3} comes 
\begin{subequations}\label{coord-changeKRP}
\begin{align}\label{proca-kalb_ramond}
    \Pi^{0 i } = - m A_i , \qquad m B_{0 i } = \dot{A}_i - \partial_i A_0,
\end{align}
and using \eqref{KR-comm} we get
\begin{align}\label{procacomm}
    [ A_i( \boldsymbol{y},t) \, , \, ( \dot{A}_j - \partial_j A_0) ( \boldsymbol{x},t) ] = i \,\delta^j_i \delta^3 ( \boldsymbol{x}- \boldsymbol{y} ).
\end{align}
\end{subequations}
These commutation relations will be satisfied as long as 
\begin{subequations}\label{defsa}
\begin{align}
   & \left[ a (\boldsymbol{p}, \sigma ) \, , \, a^{\dagger } (\boldsymbol{p}', \sigma')  \right]  = \delta_{\sigma \sigma'} \delta^3 (\boldsymbol{p} - \boldsymbol{p}' ) ; \\
   & \left[ a (\boldsymbol{p}, \sigma ) \, , \, a (\boldsymbol{p}', \sigma')  \right] = 0,
\end{align}
\end{subequations}
fixing the solution for $A_\mu(x)$. 

Using our general solution \eqref{KR3} for $B_{\mu \nu}$ and \eqref{eqmH} for the canonical momenta it is straightforward to show that \eqref{hkr} reduces to
\begin{align}\label{Hkr/proca}
    H^R_{KR} = \frac{1}{2} \sum_{\sigma} \int d^3 p \, p^0  \left( a(\boldsymbol{p} , \sigma ) a^{\dagger} (\boldsymbol{p}, \sigma)  + a^{\dagger} (\boldsymbol{p}, \sigma) a(\boldsymbol{p} , \sigma) \right). 
\end{align}
 This is just the usual free particle Hamiltonian obtained after canonical quantization of the Proca theory. The operators $a(\boldsymbol{p}, \sigma), a^{\dagger}(\boldsymbol{p}, \sigma)$ have the same properties of the creation and annihilation operators in the canonical quantization of Proca. The canonical quantization of Proca and Kalb-Ramond leads to the same free Hamiltonian and Hilbert space, in this sense the theories are dual. 

 This result can be clarified by adopting the point of view developed in \cite{weinberg_1995}, in which we analyze the canonical quantization the other way around. Starting from the Hilbert space of spin-1 particles and the free Hamiltonian
\begin{align}\label{hamilfs=1}
  H_0= \frac{1}{2} \sum_{\sigma} \int d^3 p \, p^0  a^{\dagger}(\boldsymbol{p}, \sigma ) a(\boldsymbol{p}, \sigma )  ,
\end{align}
we can implement the canonical formalism and recover the Kalb-Ramond Lagrangian by the Legendre transformation of the Hamiltonian. The reason for this Lagrangian to be associated with free spin-1 is that when canonically quantized it gives the right answer. This approach is developed in \ref{apen1}.  In \cite{weinberg_1995} this is illustrated for the Proca Lagrangian, the key point is that the Lagrangian to be recovered by Legendre transformation of the Hamiltonian depends on our choice of canonical coordinates and this choice is not unique.

\subsection{Curtright-Freund model}

Consider the Curtright-Freund vector model \cite{Curtright:1980yj}:
 \begin{align}\label{frees=0}
     S_{CF} = \int d^4 x \left[ \frac{(\partial^\mu B_\mu)^2}{2} + \frac{m^2 B_\mu B^\mu}{2} \right].
 \end{align}
 Noticing that the spatial component $\Pi^j$ of the momenta is zero, and therefore a constraint, we have 
 \begin{align}
     H_{CF} = \int  d^3 x \left[ \frac{\Pi^0 \Pi^0}{2} - \frac{m^2}{2} B_\mu B^\mu   - B_j \partial^j \Pi^0 +  \lambda_j \Pi^j \right] . 
 \end{align}
The consistency condition $\dot{\Pi}^j = 0$ leads to the new constraint
\begin{align}\label{conss=0}
    m^2 B_j + \partial^j \Pi^0  = 0,
\end{align}
whose time evolution fixes $\lambda_j$. From \eqref{dirac-brack} we note that the non-zero Dirac-brackets among independent canonical variables is
\begin{align}
     \{ B_0 (\boldsymbol{x}, t) \, , \, \Pi^0( \boldsymbol{y} ,t ) \}_D = \delta^3 (\boldsymbol{x} - \boldsymbol{y} ).
\end{align}

In order to quantize the theory we postulate the commutators
\begin{subequations}\label{comm-s=0}
\begin{align}
     & \left[ B_0 (\boldsymbol{x}, t) \, , \, \Pi^0( \boldsymbol{y} ,t ) \right] = i \delta^3 (\boldsymbol{x} - \boldsymbol{y} );\\
     & \left[ B_0 (\boldsymbol{x}, t) \, , \,B_0 (\boldsymbol{y}, t) \right]  =  \left[ \Pi^0( \boldsymbol{x} ,t ) \, , \, \Pi^0( \boldsymbol{y} ,t ) \right] = 0,
\end{align}
\end{subequations}
and the reduced Hamiltonian  
\begin{align}\label{HRs=0}
    H^R_{CF} =  \int \frac{d^3 x }{2} \left[  \Pi^0 \Pi^0  + m^2 B^2_0  + \frac{( \boldsymbol{\nabla} \Pi^0 )^2  }{ m^2}  \right].
\end{align}
is promoted to the Hamiltonian operator. The covariant equations of motion are
\begin{align}\label{eq2s=0}
    m^2 B_\mu - \partial_\mu \partial^\alpha B_\alpha = 0,
\end{align}
whose divergence gives 
\begin{align}\label{KGs=0}
    \left( \square - m^2 \right) \partial^\alpha B_\alpha = 0.
\end{align}
By the same arguments of the Kalb-Ramond case we can show that $B_\mu = \partial_\mu \phi/m$ is the general solution of \eqref{eq2s=0}, where $\phi $ is the general solution of \eqref{KGs=0}. Noticing that 
\begin{subequations}\label{coordchangeCF-KG}
\begin{align}
    B_0  = \frac{\dot{\phi}}{m}, \qquad \qquad \Pi^0 = - m \phi
\end{align}
and using \eqref{comm-s=0} we can show that 
\begin{align}\label{comms=0}
   & \left[ \phi( \boldsymbol{x}, t ) \, , \, \dot{\phi} ( \boldsymbol{y} ,t ) \right]  =  i \delta^3 ( \boldsymbol{x} - \boldsymbol{y} ),\\ \label{comm-s=02}
   & \left[ \phi (\boldsymbol{x}, t) \, , \,\phi (\boldsymbol{y}, t) \right]  =  \left[ \dot{\phi}( \boldsymbol{x} ,t ) \, , \, \dot{\phi}( \boldsymbol{y} ,t ) \right] = 0.
\end{align}
\end{subequations}
The general solution of \eqref{KGs=0} has the Fourier expansion 
\begin{align}
    \phi (\boldsymbol{x},t) = (2 \pi)^{-3/2}  \int
    \frac{d^3 p }{(2 p^0)^{1/2}} \left(  a( \boldsymbol{p}) e^{i p \cdot x} 
      +   a^{\dagger} (\boldsymbol{p}) e^{-i p \cdot x} \right).
\end{align}
Using \eqref{comms=0} and \eqref{comm-s=02} we obtain 
\begin{subequations}
\begin{align}
     & \left[ a (\boldsymbol{p} ) \, , \, a^{\dagger } (\boldsymbol{p}')  \right]  =  \delta^3 (\boldsymbol{p} - \boldsymbol{p}' ) ; \\
   & \left[ a (\boldsymbol{p}) \, , \, a (\boldsymbol{p}')  \right] = 0.
\end{align}
\end{subequations}
In terms of $B_\mu = \partial_\mu \phi/m$ \eqref{HRs=0} is just  
\begin{align}\label{HCF-KG}
    H^R_{CF} = \frac{1}{2} \int d^3 p \, p^0  \left( a(\boldsymbol{p}) a^{\dagger} (\boldsymbol{p})  + a^{\dagger} (\boldsymbol{p}) a(\boldsymbol{p}) \right). 
\end{align}
Therefore the Curtright-Freund vector model is dual to Klein-Gordon in the same sense that Kalb-Ramond is dual to Proca. As one may expect the same argument of \ref{apen1} may be used for the spin-0 case, the development is found in \ref{apens=0}.

As illustrated by \eqref{coord-changeKRP}, \eqref{Hkr/proca}, \eqref{coordchangeCF-KG} and \eqref{HCF-KG} the equivalence between massive dual models in the canonical formalism translates to a canonical coordinate change: the dual Lagrangians derive the same Hamiltonian in different canonical coordinates. This relation between free theories is crucial to define the duality of Feynman rules for interacting massive dual models in the canonical formalism.

  \section{Interaction-picture}\label{interaction-picture2}
 
  We have seen that the free Kalb-Ramond field is a function of the free Proca field and the vector field of spin-0 is a function of the Klein-Gordon field. This is relevant when it comes to perturbation theory, to see why we shall review how perturbation theory works. Given an Hamiltonian $H$ as a functional of fields $\Psi_l(\boldsymbol{x} , t  ), \Pi_l(\boldsymbol{x}, t )$, in the Heisenberg picture, the fields in the interaction picture are given by the similarity transformation 
  \begin{align}\label{V(t)}
      \Psi^I_l (\boldsymbol{x},t) =  e^{i H_0 t} \Psi_l (\boldsymbol{x},0) e^{- i H_0  t} \\
      \Pi^I_l (\boldsymbol{x},t) = e^{i H_0 t} \Pi_l (\boldsymbol{x},0) e^{- i H_0  t}
  \end{align}
 where $H_0 $ is the free Hamiltonian and the upper index $I$ stands for interaction  picture. This similarity transformation on $H$ gives
 \begin{align}
     e^{i H_0 t} H e^{- i H_0  t} = e^{i H_0 t} (H_0  +  V) e^{- i H_0  t} = H_0 + V(t),
 \end{align}
 where $V$ is the interacting part of the full Hamiltonian $H$. The term $V(t)$ is expressed as an integral of the Hamiltonian density $\mathcal{H}(x)$:
 \begin{align}
     V(t) = \int d^3x \, \mathcal{H}(x),
 \end{align}
 so that the S-matrix~\cite{Wheeler:1937zz,Heisenberg1,Heisenberg2} is given by the power series\cite{PhysRev.75.1736}
 \begin{align}
     S  =  \mathbf{1}  + \sum_{n=1}^{\infty} \frac{(-i)^n}{n!} \int d^4 x_1 \cdots d^4 x_n T\{\mathcal{H}(x_1) \cdots \mathcal{H}(x_n)\},
 \end{align} 
 where $T$ is the time ordering product. The interaction Hamiltonian $\mathcal{H}(x)$ is a function  of the free fields $\Psi^I_l$ and $\Pi^I_l$. Expressing $\Pi^I_l$ as a function of $\Psi^I_l$ we rewrite $\mathcal{H}(x)$ in terms of $\Psi^I_l$, and its derivatives. An important quantity in perturbation theory is the propagator of the fields $\Psi^I_l$ in $\mathcal{H}(x)$:
 \begin{align}\label{propdef}
     - i \Delta_{lm} (x,y) \equiv \langle T \{ \Psi^I_l(x) \Psi^I_{m} (y) \} \rangle_0.
 \end{align}
 
 Since the free Kalb-Ramond field is a function of the free Proca field any $\mathcal{H}(x)$ written in terms of $B_{0i} \, , \, \Pi^{0i}$ could in principle be rewritten in terms of $A_j \, , \, \pi^j$, where $\pi^j$ is the canonical momenta of the Proca theory given in \eqref{proca-kalb_ramond} in terms of $B_{0j}$. One would expect that, perturbatively, any interacting Kalb-Ramond Lagrangian is equivalent to some interacting Proca Lagrangian, in the sense that they generate the same $\mathcal{H}(x)$. Naturally, we can use the same argument to propose an equivalence of Feynman rules derived from the Curtright-Freund model and Klein-Gordon. The equivalence between Feynman rules can also be formulated in terms of the path integral, where its generality is clearer. In the next section, we propose an interacting Kalb-Ramond Lagrangian. The interaction will be worked in the canonical and Path Integral formalism. 
 
 \section{Interacting spin-1}

  Consider Kalb-Ramond interacting with a massless scalar field through a "Higgs-like" interaction \begin{align}\label{interacting-1}
     S = \int d^4 x \left[ - \frac{1}{2} \partial_\mu  \phi \partial^\mu \phi + \frac{1}{2} \partial_\alpha B^{\alpha \mu} \partial^{\nu} B_{\nu \mu} + \frac{m^2}{4} B_{\mu \nu} B^{\mu \nu} + \frac{\lambda}{4} \phi^2 B_{\mu \nu} B^{\mu \nu} \right],
  \end{align}
  with $\lambda >0$. The theory of the Higgs mechanism when the bare mass is not zero has been considered in  \cite{Konoue:1976pz}. 
The conjugate momenta are 
\begin{align}
    \pi = \dot{\phi}, \qquad \Pi^{0 i } = - \partial_{\alpha} B^{\alpha i}, \qquad \Pi^{ij} = 0,
\end{align}
and the Hamiltonian is 
\begin{align}
H = \int d^3 x  \bigg[ \frac{1}{2} \pi^2 + \frac{1}{2} \Pi^{0 i } \Pi^{0 i } + \Pi^{0 i } \partial^j B_{j i } + \frac{1}{2} (\boldsymbol{\nabla} \phi )^2  + \frac{1}{2} (\partial^i B_{0i})^2 + \Lambda_{ij} \Pi^{ij} \nonumber  \\ -  \frac{1}{4} (m^2 + \lambda \phi^2) B_{\mu \nu} B^{\mu \nu}  
   \bigg].
\end{align}
The time derivative $\dot{\Pi}^{ij}=0$ now gives 
\begin{align}\label{B_ij}
    \partial^i \Pi^{0j} - \partial^{j} \Pi^{0i}  + m^2 \left(1 + \frac{\lambda \phi^2}{m^2}\right) B_{ij}  = 0,
\end{align}
whose time derivative fixes $\Lambda_{ij}$. We can already see the first interesting aspect  of interacting Kalb-Ramond: the constraint equation is modified in such a way that $B_{ij}$ becomes a non-linear function of $\phi$
\begin{align}
    B_{ij}  = - \frac{1}{m^2} \left(1 + \frac{\lambda \phi^2}{m^2}\right)^{-1} \left( \partial^i \Pi^{0j} - \partial^{j} \Pi^{0i} \right).
\end{align}
The consequence is that the reduced Hamiltonian is also a non-linear function of $\phi$
\begin{align}\label{fullHInKR}
H^R_{KR} = \int d^3 x  \bigg[ \frac{1}{2} \pi^2 + \frac{1}{2} \Pi^{0 i } \Pi^{0 i } +  \frac{1}{2} (\boldsymbol{\nabla} \phi )^2  + \frac{1}{2} (\partial^i B_{0i})^2 +  \frac{m^2}{2} \left( 1 + \frac{\lambda \phi^2}{m^2}\right) B_{0 i} B_{0 i} \nonumber\\ + \frac{1}{2 m^2} \left(1 + \frac{\lambda \phi^2}{m^2}\right)^{-1} ( \boldsymbol{\nabla} \times \boldsymbol{\Pi} )^2
\bigg].
\end{align}
Since $\phi$ and $\Pi^{0j}$ have zero Poisson brackets with all constraints, the non-zero commutators among independent variables coming from Dirac brackets are 
\begin{align}
    [ \phi \, , \, \pi ] =  i \,  \delta^3 (\boldsymbol{x} - \boldsymbol{y}  ), \qquad [B_{0i}  \, , \, \Pi^{0 j} ]  = i \, \delta^i_j \delta^3 (\boldsymbol{x} - \boldsymbol{y} ),  
\end{align}

Up to now we have been working in the Heisenberg-picture, where the fields evolve with the full Hamiltonian \eqref{fullHInKR}. The interaction Hamiltonian reads
\begin{align}
    \mathcal{H}_{NC}(x) =   \frac{\lambda \phi^2}{2} B_{0i} B_{0i} + \frac{1}{2m^2} ( \boldsymbol{\nabla} \times \boldsymbol{\Pi} )^2 \left(  \sum_{n=1}^{\infty} (-1)^n \left(\frac{\lambda \phi^2}{m^2} \right)^n \right) ,
\end{align}
with the fields in the interaction picture. Since the fields are free  we can use \eqref{proca-kalb_ramond} to write 
\begin{align}\label{V(t)IntKR}
     \mathcal{H}_{NC}(x) =   \frac{\lambda \phi^2}{2 m^2} F_{0i} F_{0i} + \frac{1}{4} \left( \sum^{\infty}_{n=1} (-1)^n \left(\frac{\lambda \phi^2}{m^2} \right)^{n} \right) F_{ij} F_{ij} .
\end{align}
 The theory is highly nonlinear in the scalar field, which is a consequence of the modification \eqref{B_ij} in the constraint equation.
 
  The term \eqref{V(t)IntKR} can also be derived from the action
 \begin{align}\label{interacting-2}
S = \int d^4 x \left[- \frac{1}{4} \left( 1 + \frac{\lambda \phi^2}{m^2 } \right)^{-1} F_{\mu \nu} F^{\mu \nu} - \frac{m^2}{2} A_\mu A^\mu - \frac{1}{2} \partial_\mu \phi \partial^\mu\phi \right].
\end{align}
 The only primary constraint now is the zero component of the Proca momentum $\bar{\Pi}^0 = 0$. The Hamiltonian reads
\begin{align}
    H = \int d^3 x \Bigg[ \left( 1 + \frac{\lambda \phi^2 }{m^2 } \right) \frac{\bar{\boldsymbol{\Pi}}^2}{2} + \frac{\pi^2}{2} - A_0 \boldsymbol{\nabla} \cdot \bar{\boldsymbol{\Pi}} + \left( 1 + \frac{\lambda \phi^2 }{m^2 } \right)^{-1} \frac{(\boldsymbol{\nabla} \times \boldsymbol{A} )^2 }{2} + \frac{(\nabla \phi)^2}{2} \nonumber\\
    + \frac{m^2 }{2} A_\mu A^\mu + \bar{\lambda} \bar{\Pi}^0 \Bigg].
\end{align}
The time evolution $\dot{\bar{\Pi}}^0=0$ leads to the new constraint 
\begin{align}
    m^2 A_0 + \boldsymbol{\nabla} \cdot \bar{\boldsymbol{\Pi}} =  0,
\end{align}
whose time evolution fixes $\bar{\lambda}$. The reduced Hamiltonian is 
\begin{align}\label{reducedproca}
    H^R_{P} = \int \frac{d^3 x }{2 } \Bigg[ \left( 1 + \frac{\lambda \phi^2 }{m^2 } \right) \bar{\boldsymbol{\Pi}}^2 + \pi^2 + \left( 1 + \frac{\lambda \phi^2 }{m^2 } \right)^{-1} (\boldsymbol{\nabla} \times \boldsymbol{A} )^2  + (\boldsymbol{\nabla} \phi)^2 \nonumber\\
    + m^2 \boldsymbol{A}^2 + \frac{(\boldsymbol{\nabla} \cdot \bar{\boldsymbol{\Pi}})^2}{m^2} \Bigg].
\end{align}
The non-zero Dirac-brackets among independent canonical variables are
\begin{align}
    \{ A_i( \boldsymbol{x},t) \, , \, \bar{\Pi}^j (\boldsymbol{y}, t) \}_D = \delta^j_i \, \delta^3(\boldsymbol{x}- \boldsymbol{y} ) \, , \qquad \{ \phi(\boldsymbol{x},t) \, , \, \pi(\boldsymbol{y},t) \}_D = \delta^3(\boldsymbol{x}- \boldsymbol{y} ),
\end{align}
then to quantize the theory we postulate the only non-zero commutation relations of independent canonical variables:
\begin{align}
    \left[ A_i( \boldsymbol{x},t) \, , \, \bar{\Pi}^j (\boldsymbol{y}, t) \right] = \delta^j_i \, \delta^3(\boldsymbol{x}- \boldsymbol{y} ) \, , \qquad \left[ \phi(\boldsymbol{x},t) \, , \, \pi(\boldsymbol{y},t) \right] = \delta^3(\boldsymbol{x}- \boldsymbol{y} )
\end{align}

From \eqref{reducedproca} we obtain the interaction Hamiltonian
\begin{align}
   \mathcal{H}_{NC}(x) = \ \frac{\lambda \phi^2 \bar{\boldsymbol{\Pi}}^2}{2 m^2 } + \left( \sum^{\infty}_{n=1} (-1)^n \left(\frac{\lambda \phi^2}{m^2} \right)^{n} \right) \frac{(\boldsymbol{\nabla} \times \boldsymbol{A} )^2}{2} .
\end{align}
Since these fields are free we can use $\bar{\Pi}^i = - F^{0i}$, so that 
\begin{align}
     \mathcal{H}_{NC}(x) =  \frac{\lambda \phi^2}{2 m^2} F_{0i} F_{0i} + \frac{1}{4} \left( \sum^{\infty}_{n=1} (-1)^n \left(\frac{\lambda \phi^2}{m^2} \right)^{n} \right) F_{ij} F_{ij} ,
\end{align}
this is the same term \eqref{V(t)IntKR} obtained from \eqref{interacting-1}.

The existence of an interacting Lagrangian with the Proca field $A_\mu$ and the same perturbative content was already expected by the arguments of the last section. The subtle point is that a simple Lagrangian like \eqref{interacting-1} is related to a Lagrangian with nonlinear terms of all orders like \eqref{interacting-2}. This is expected to be the case whenever the constraint equation \eqref{xi_ij} is modified in the following sense:
\begin{align}\label{constraintchange}
   \partial^i \Pi^{0j} -\partial^j \Pi^{0i} +  m^2 B_{ij}  = 0 \rightarrow \partial^i \Pi^{0j} -\partial^j \Pi^{0i} +  \left( \delta^l_i \delta^k_j + M^{lk}_{ij} \right) B_{lk} = 0,
\end{align}
with $M^{lk}_{ij}$ being a function of the fields  with which we couple $B_{\mu \nu}$. For a non zero $M^{lk}_{ij}$ the solution of $B_{i  j}$ in terms of the independent canonical fields will be a geometric series in $M^{lk}_{ij}$, leading to a highly nonlinear Hamiltonian, since we always have a mass term $m^2B^2_{ij}$.  In the scalar interaction example $M^{lk}_{ij}$ is proportional to a product of deltas. However, if we introduce couplings with tensor fields $M^{lk}_{ij}$ is expected to be a more complicated function.

\subsection{Covariant Feynman rules}

The interaction Hamiltonian
 \begin{align}\label{v(t)}
     \mathcal{H}_{NC}(x) =   \frac{\lambda \phi^2}{2 m^2} F_{0i} F_{0i} + \frac{1}{4} \left( \sum^{\infty}_{n=1} (-1)^n \left(\frac{\lambda \phi^2}{m^2} \right)^{n} \right) F_{ij} F_{ij} ,
 \end{align}
 is not covariant. Non covariant interaction Hamiltonians are an old problem in canonical perturbation theory \cite{weinberg_1995,Greiner:1996zu,Rohrlich:1950zz,PhysRev.76.684.2}.
 The interaction is expected to be non-covariant whenever we have spin $s \geq 1$, field derivatives, or both in the interaction lagrangian.  Usually, the non-covariant terms in the Hamiltonian density cancel the non-covariant part of \eqref{propdef} so that one can do perturbation theory with a covariant Hamiltonian density and covariant propagators. However, that is not always the case \cite{Gerstein:1971fm,Lee:1962vm}. 
 
 With the time-like vector $n_\mu  = (1,0,0,0)$ we can rewrite \eqref{v(t)}:
 \begin{subequations}
 \begin{align}\label{FOF}
      \mathcal{H}_{NC}(x)= -& \frac{1}{2}  F^{\mu \nu} O_{\mu \nu \alpha \beta} F^{\alpha \beta}, \\
      O_{\mu \nu \alpha \beta } \equiv
      - \frac{(\lambda \phi^2 /m^2)^2}{4\left(1+\lambda \phi^2 /m^2 \right)} &  \left( n_\mu n_\alpha \eta_{\nu \beta} - n_\nu n_\alpha \eta_{\mu \beta} - n_\mu n_\beta \eta_{\nu \alpha} + n_\nu n_\beta \eta_{\mu \alpha} \right) \nonumber\\
      &+ \frac{\lambda \phi^2}{4 m^2} \left( 1 - \frac{\lambda \phi^2/m^2}{1+ \lambda \phi^2 /m^2} \right) \left( \eta_{\alpha \mu} \eta_{\beta \nu} - \eta_{\beta \mu} \eta_{\alpha \nu} \right).
      \end{align}
 \end{subequations}
 The field propagators are \begin{subequations}
 \begin{align}\label{phim=0prop}
     &\langle T\{ \phi(x) \phi (y) \} \rangle_0 = - i \,  \Delta_F^{m=0}(x,y), \\ \label{FFprop}
    & \langle T \{ F^{\mu \nu}(x) F^{\alpha \beta}(y) \} \rangle_0  = - i \, \Bigg[
     \delta^4 (x-y) \left(\delta^\mu_0 \delta^\alpha_0 \eta^{\nu \beta} - \delta^\mu_0 \delta^\beta_0 \eta^{\nu \alpha} - \delta^\nu_0 \delta^\alpha _0 \eta^{\mu \beta} + \delta^\nu_0 \delta^\beta_0 \eta^{\mu \alpha} \right) \nonumber\\
     &-\partial^\mu \partial^\alpha \Delta^{\nu \beta}(x,y) + \partial^{\mu} \partial^\beta \Delta^{\nu \alpha} (x,y) + \partial^\nu \partial^\alpha \Delta^{\mu \beta} (x,y)  - \partial^\nu \partial^\beta  \Delta^{\mu \alpha}(x,y) \Bigg]  ,\\
     &\Delta^{\nu \beta}(x,y) \equiv \left( \eta^{\mu \nu} - \frac{\partial^\mu \partial^\nu }{m^2} \right) \Delta_F (x,y) ,
 \end{align}
 \end{subequations}
where $\Delta_F^{m=0}(x,y)$ and $\Delta_F(x,y)$ are the Feynman deltas for zero and non-zero mass respectively. Our present problem is to sum the contributions of the non-covariant contact term in \eqref{FFprop} so that we end up with a covariant perturbation theory. We note that \eqref{FOF} has the same form worked out in \cite{Lee:1962vm}, therefore we can use the same diagrammatic technique.

 Following \cite{Lee:1962vm} we represent \eqref{FFprop} by a sum of straight and spring lines, where the spring lines represent the non-covariant part of the propagator. The effect of inserting $n-1$ spring lines between two straight lines is the contribution 
 \begin{align}
    I_n &= i \,  \frac{4^{n-1}}{2} \int d^4x \, F^{\alpha \beta} F^{\mu \nu} O_{\alpha \beta 0 i} O_{0 i 0 j} \cdots O_{0 p \kappa \xi }\nonumber\\
    &= \frac{i}{2} \int d^4 x F^{0 i } F^{0 i } \left( - \frac{\lambda \phi^2}{m^2} \right)^n, \qquad n\geq 2
\end{align}
 The sum over all $n$, including n=1 which is just \eqref{v(t)} gives the corrected vertex contribution
 \begin{align}
   (-i) \int d^4 x \left( - \frac{\lambda \phi^2 }{4m^2} \frac{F_{\mu \nu}F^{\mu \nu} }{\left(1 + \lambda \phi^2 /m^2 \right)} \right).
 \end{align}
  The spring lines can also form closed loops, the contribution of a loop with $n$ springs is 
\begin{align}
      \bar{I}_n = - \frac{3}{2} \delta^4 (0) \int d^4 x \frac{(-1)^{n-1}}{n} \left(\frac{\lambda \phi^2}{n} \right), \qquad n \geq 1,
  \end{align}
the sum over all $n $ gives the contribution 
\begin{align}
    - \frac{3}{2} \delta^4(0) \int d^4 x \, \ln \left(1+ \lambda \phi^2/m^2 \right).
\end{align}
The net result is that we can use 
\begin{align}
    \mathcal{H}_{C}(x) =   - \frac{\lambda \phi^2 }{4m^2} \frac{F_{\mu \nu}F^{\mu \nu} }{\left(1 + \lambda \phi^2 /m^2 \right)} - \frac{3 \, i }{2} \delta^4(0) \ln \left( 1 + \lambda \phi^2/m^2 \right)  
\end{align}
with the covariant part of \eqref{FFprop} in perturbation theory calculations. 

Starting with either \eqref{interacting-1} or \eqref{interacting-2} we get the same Hamiltonian density \eqref{v(t)} however, the usual drawback is that one has to sum the non-covariant propagator contributions to obtain the covariant perturbation theory. The cure for this inconvenience is the path integral formulation, where the Feynman rules are covariant from the start. However, in the path integral the perturbative equivalence of \eqref{interacting-1} and \eqref{interacting-2} is not manifest and one has to sum contact terms to show that they define the same perturbation theory. 

\subsection{Path integral Feynman rules}

 The Lagrangian path integral associated to \eqref{interacting-1} and \eqref{interacting-2} are respectively
\begin{subequations}\ref{PI}
\begin{align}\label{PI_KR}
     &\langle Vac,in |  Vac,out \rangle =
     \mathcal{C} \int \mathcal{D} \phi \mathcal{D} B_{[\mu \nu]} \, e^{   i \int d^4 x \left[ \mathcal{L}^0_{KR}
    + \frac{\lambda}{4} \phi^2 B_{\mu \nu} B^{\mu \nu}  -\frac{3 \, i }{2} \delta^4(0) \ln \left(1+ \lambda \phi^2 /m^2 \right)\right]  }, \\ \label{PI_P}
    &\langle Vac,in |  Vac,out \rangle = \mathcal{C}' \int \mathcal{D} \phi \mathcal{D} A_{\mu} \, e^{   i \int d^4 x \left[ \mathcal{L}^0_{P}
    + \frac{\lambda \phi^2}{4 m^2}  \frac{F_{\mu \nu} F^{\mu \nu}}{\left(1+ \lambda \phi^2/m^2\right)}  + \frac{3 \, i }{2} \delta^4(0) \ln \left(1+ \lambda \phi^2 /m^2 \right)\right]    },
\end{align}
where $\mathcal{C}, \mathcal{C}'$ are field independent constants and
\begin{align}\label{kineticB}
    \mathcal{L}^0_{KR} &\equiv - \frac{1}{2} \partial_\mu  \phi \partial^\mu \phi + \frac{1}{2} \partial_\alpha B^{\alpha \mu} \partial^{\nu} B_{\nu \mu} + \frac{m^2}{4} B_{\mu \nu} B^{\mu \nu} \\
    \mathcal{L}^0_P &\equiv - \frac{1}{2} \partial_\mu \phi \partial^\mu\phi - \frac{1}{4} F_{\mu \nu} F^{\mu \nu} - \frac{m^2}{2} A_\mu A^\mu .
\end{align}
\end{subequations}
The perturbation theory derived from \eqref{PI_P} is equal to the canonical formalism one of the last section. However, is not obvious that the perturbation theory derived from \eqref{PI_KR} is equivalent to the one coming from \eqref{PI_P}.

In order to prove the equivalence we recast the problem in the language of the canonical formalism. The Feynman rules coming from \eqref{PI_KR} are the same that one would derive from 
\begin{align}\label{H_PI}
\mathcal{H}_{PI}(x) =  -\frac{\lambda \phi^2}{4} B_{\mu \nu} B^{\mu \nu} + \frac{3 \, i }{2} \delta^4(0) \ln \left( 1 + \lambda \phi^2 /m^2 \right) ,
\end{align}
with the $B_{\mu \nu}(x)$ propagators 
\begin{align}\label{PIpropB}
    &\langle T \{ B^{\mu \nu}(x) B^{\alpha \beta}(y) \} \rangle_0  = - i \Bigg[
     - \frac{\delta^4 (x-y)}{m^2} \left(\eta^{\mu \alpha} \eta^{\nu \beta} - \eta^{\mu \beta} \eta^{\nu \alpha} \right) \nonumber\\
     &+ \frac{1}{m^2} \left[-\partial^\mu \partial^\alpha \Delta^{\nu \beta}(x,y) + \partial^{\mu} \partial^\beta \Delta^{\nu \alpha} (x,y) + \partial^\nu \partial^\alpha \Delta^{\mu \beta} (x,y)  - \partial^\nu \partial^\beta  \Delta^{\mu \alpha}(x,y)  \right] \Bigg]
\end{align}
and \eqref{phim=0prop} as the $\phi(x)$ propagator. The path integral propagator \eqref{PIpropB} is defined as the inverse of the kinetic term in \eqref{kineticB}. 

We can sum the effect of the contact terms in \eqref{PIpropB} with the same technique of the last section. Inserting $n-1$ spring lines between to straight lines, and then summing from $n=1$ to $n= \infty$ gives the corrected vertex 
\begin{align}
    - i \int d^4 x \left[ - \frac{\lambda \phi^2 B_{\mu \nu} B^{\mu \nu} }{4\left(1+\lambda \phi^2/m^2\right) } \right].
\end{align}
Adding up the contributions of loops formed by spring lines gives 
\begin{align}
- 3 \delta^4(0) \int d^4 x \ln \left( 1 + \lambda \phi^2 /m^2 \right).
\end{align}
The effect of taking this corrections into account is that \eqref{H_PI} is replaced by the effective Hamiltonian density
\begin{align}
\mathcal{H}_{eff}(x) = - \frac{\lambda \phi^2 B_{\mu \nu} B^{\mu \nu} }{4\left(1+\lambda \phi^2/m^2\right) } - \frac{3 \, i }{2} \delta^4(0) \ln \left( 1 + \lambda \phi^2 /m^2 \right),
\end{align}
and the propagator for $B_{\mu \nu}$ is given by the second line in \eqref{PIpropB},
that is just the covariant part of  \eqref{FFprop} multiplied by $m^{-2}$. The difference in the Feynman rules derived from \eqref{PI_KR} and \eqref{PI_P} is a consequence of the contact term in \eqref{PIpropB}. Adding up the effect of this contact term allows us to recover the Feynman rules of \eqref{PI_P} from \eqref{PI_KR}.

Our specific choice of interaction should not deviate our attention from general points that can be made. Given any interacting Kalb-Ramond Lagrangian $\mathcal{L}^I$ in the path integral, the propagator of the Kalb-Ramond field is always \eqref{PIpropB}, since it is derived from the free Lagrangian. Therefore, any set of Feynman rules derived from a Kalb-Ramond Lagrangian will initially contain contact terms, referent to the Kalb-Ramond fields in $\mathcal{L}^I$. In order to redefine the Feynman rules, we summed a subset of all possible diagrams, those including contact terms. The technique we used to do the sum works for one type of interaction, for other types one should use a different technique, but the sum should always be done. The Feynman rules after the redefinition will be generated by an effective interaction Lagrangian $\mathcal{L}^I_{eff}$ depending on the Kalb-Ramond fields. With this new vertex, the propagators between Kalb-Ramond fields will be just the second line of \eqref{PIpropB}. These are the covariant Feynman rules that one would derive from the path integral by replacing the free Kalb-Ramond Lagrangian with the free Proca Lagrangian and the Kalb-Ramond fields in $\mathcal{L}^I_{eff}$ by $F_{\mu \nu}/m$. This is the path integral version of our argument in \ref{interaction-picture2} for the equivalence between the Feynman rules of dual massive models.

The path integral propagator of $A_\mu$ in the Curtright-Freund model is
\begin{align}
    \langle T \{ A_\mu (x) A_\nu(y) \} \rangle_{PI}  =  \frac{- i }{m^2} \left[ - \partial_\mu \partial_\nu \Delta(x,y) - \eta_{\mu \nu} \delta^4(x-y) \right],
\end{align}
the first term is just the covariant propagator of $\partial_\mu \phi$ multiplied by $m^{-2}$. It is straightforward to extend the arguments of Kalb-Ramond to the spin-0 vector model and Klein-Gordon. In the context of path integral the Feynman rules of Kalb-Ramond and Curtright-Freund are equivalent to Proca and Klein-Gordon once we sum the effect of the contact terms.

    \section{Beyond perturbation theory}
    
     We have shown in \ref{apen1} how the free Hamiltonians of Kalb-Ramond and Proca are connected by the canonical transformations 
    \begin{align}\label{canKR-P}
        q_i (\boldsymbol{x},t) = m^{-1} \pi_P^i (\boldsymbol{x},t), \qquad \pi^i(\boldsymbol{x},t) = - m \, q_i^P(\boldsymbol{x},t),
    \end{align}
    where $\left(\pi_P^i (\boldsymbol{x},t) \, , \, q_i^P(\boldsymbol{x},t)\right)$ are the canonical variables of Proca. With the identification 
    \begin{align}
       &\boldsymbol{q}^P(\boldsymbol{x},t) \equiv \boldsymbol{A}(\boldsymbol{x},t), \qquad \boldsymbol{\pi}_P(\boldsymbol{x},t) \equiv \bar{\boldsymbol{\Pi}}(\boldsymbol{x},t);\\
        &q_i(\boldsymbol{x},t) \equiv B_{0i}(\boldsymbol{x},t), \qquad \pi^i(\boldsymbol{x},t) \equiv \Pi^{0i}(\boldsymbol{x},t),
    \end{align}
    it is easy to see that \eqref{canKR-P} also connects the interacting Hamiltonians \eqref{fullHInKR} and \eqref{reducedproca}. These fields are in the Heisenberg picture, therefore the theories are not only the same in the perturbative sense, they are the same even non-perturbatively. Given any quantum Hamiltonian coming from an interacting Proca Lagrangian, we can always perform the coordinate change  \eqref{canKR-P}. The non trivial point is whether this coordinate change will lead to a Hamiltonian that can be derived from a Kalb-Ramond Lagrangian. In our simple example that is the case and as we are going to see this may be true for more general interactions.
    
    Despite having shown that \eqref{interacting-1} and \eqref{interacting-2} define the same physics, we did not mention how we knew beforehand that these two Lagrangians would have the same content. This was guessed from the parent action method \cite{Deser:1984kw}. 
    
    Consider 
  \begin{align}\label{SpKR-P}
         S_P  = \int d^4 x \Bigg[ - \frac{m^2}{2}A_\mu A^\mu + \frac{m^2}{4} B_{\mu \nu} B^{\mu \nu} - m B^{\mu \nu} \partial_\mu A_\nu + \frac{\lambda \phi^2 }{4}  B_{\mu \nu} B^{\mu \nu} \nonumber\\
         -\frac{\partial_\mu \phi \partial^\mu \phi}{2} \Bigg], 
     \end{align}
    If we do the coordinate change 
  \begin{align}\label{coorchaange1}
        B_{\mu \nu} \rightarrow \bar{B}_{\mu \nu} = B_{\mu \nu} -  \left( 1 + \frac{\lambda \phi^2}{m^2} \right)^{-1} \frac{F_{\mu \nu}(A)}{m},
    \end{align}
    the action \eqref{SpKR-P} reads
    \begin{align}
        S_P = \int d^4 x \Bigg[- \frac{1}{4} \left( 1 + \frac{\lambda \phi^2}{m^2 } \right)^{-1} F_{\mu \nu} F^{\mu \nu} - \frac{m^2}{2} A_\mu A^\mu - \frac{1}{2} \partial_\mu \phi \partial^\mu\phi \nonumber\\
       + \frac{m^2}{4} \left(1 + \frac{\lambda \phi^2}{m^2} \right) \bar{B}_{\mu \nu } \bar{B}^{\mu \nu} \Bigg].
    \end{align}
    We can drop the term $\bar{B}^2_{\mu \nu}$ since its equations of motion are $\bar{B}_{\mu \nu} = 0$\footnote{This is also consistent with the canonical formalism because the absence of derivatives generates the primary contraints $\bar{\Pi}^{\mu \nu} = 0 $, whose time evolution gives $\bar{B}_{\mu \nu} = 0$, so the dropping of $\bar{B}_{\mu \nu}$ does not affect the quantum structure.} and work with \begin{align}
        S_P = \int d^4 x \Bigg[- \frac{1}{4} \left( 1 + \frac{\lambda \phi^2}{m^2 } \right)^{-1} F_{\mu \nu} F^{\mu \nu} - \frac{m^2}{2} A_\mu A^\mu - \frac{1}{2} \partial_\mu \phi \partial^\mu\phi  \Bigg],
    \end{align}
    which is just \eqref{interacting-2}. On the other hand if we integrate \eqref{SpKR-P} by parts, so that \begin{align}
         S_P  = \int d^4 x \Bigg[ - \frac{m^2}{2}A_\mu A^\mu + \frac{m^2}{4} B_{\mu \nu} B^{\mu \nu} + m \partial_\mu B^{\mu \nu}  A_\nu + \frac{\lambda \phi^2 }{4}  B_{\mu \nu} B^{\mu \nu} \nonumber\\
         -\frac{\partial_\mu \phi \partial^\mu \phi}{2} \Bigg], 
     \end{align}
    and then make the change of coordinates
\begin{align}\label{coorchaange2}
        A_\mu \rightarrow \bar{A}_\mu = A_\mu - \frac{\partial^\nu B_{\mu \nu}}{m}
    \end{align}
    the action now is 
    \begin{align}
         S_P = \int d^4 x \Bigg[ - \frac{1}{2} \partial_\mu  \phi \partial^\mu \phi + \frac{1}{2} \partial_\alpha B^{\alpha \mu} \partial^{\nu} B_{\nu \mu} + \frac{m^2}{4} B_{\mu \nu} B^{\mu \nu} + \frac{\lambda}{4} \phi^2 B_{\mu \nu} B^{\mu \nu} \nonumber\\
         - \frac{m^2}{2} \bar{A}_\mu \bar{A}^\mu \Bigg],
    \end{align}
    which is the same as \eqref{interacting-1} dropping $\bar{A}^2_\mu$. Therefore, \eqref{interacting-1} and \eqref{interacting-2} can be obtained from \eqref{SpKR-P}. In order to obtain \eqref{interacting-1} and \eqref{interacting-2} we had to integrate by parts, this is not a problem: the canonical quantum theory is not affected by this integration by parts \cite{weinberg_1995}. Moreover, the quantum theory is also not affected by the drop of quadratic terms like $\bar{A}_\mu^2$ and $\bar{B}^2_{\mu \nu}$. The remaining steps are the coordinate changes \eqref{coorchaange1} and \eqref{coorchaange2}. These coordinate changes are invertible, so in principle one would expect that they do not affect the physics. We checked that this is true for the free theories and simple interaction proposed here: the same Hamiltonian is derived from \eqref{interacting-1} and \eqref{interacting-2}, it is just expressed in different canonical coordinates. If we consider interactions linear in the Kalb-Ramond field, it is straightforward to propose the parent action
      \begin{align}\label{SPlinear}
         S_P  = \int d^4 x \Bigg[ - \frac{m^2}{2}A_\mu A^\mu + \frac{m^2}{4} B_{\mu \nu} B^{\mu \nu} - m B^{\mu \nu} \partial_\mu A_\nu + B_{\mu \nu} J^{\mu \nu}( \Psi_l )
        + \mathcal{L}(\psi_l) \Bigg], 
     \end{align}
    where $\Psi_l$ are not the Kalb-Ramond or Proca fields, $J^{\mu \nu}$ is a function of $\Psi_l$ and its derivatives and $\mathcal{L}(\Psi_l)$ is the free Lagrangian of the matter fields. If we integrate by parts and do the coordinate change \eqref{coorchaange2} we get Kalb-Ramond linearly coupled to the matter fields. On the other hand, performing 
\begin{align}
     B_{\mu \nu} \rightarrow \bar{B}_{\mu \nu} = B_{\mu \nu} -  \frac{F_{\mu \nu}(A)}{m} + \frac{2 J_{\mu \nu} }{m^2}
\end{align}
we get  
\begin{align}
        S_P = \int d^4 x \Bigg[- \frac{F_{\mu \nu} F^{\mu \nu}}{4}   - \frac{m^2}{2} A_\mu A^\mu + \frac{F^{\mu \nu} J_{\mu \nu} }{m}  - \frac{J_{\mu \nu} J^{\mu \nu} }{m^2} + \mathcal{L}(\psi_l) 
       + \frac{m^2}{4} \bar{B}_{\mu \nu } \bar{B}^{\mu \nu} \Bigg].
\end{align}
    We can not check explicitly whether the Lagrangians derived from the Parent action \eqref{SPlinear} derive the same Hamiltonian in different canonical coordinates, since the canonical structure depends on the matter fields and the interaction, but it seems reasonable that this should be the case. We can easily extend these arguments to the Curtright-Freund model. 
    
    The usual approach to the parent action is to take it directly in the functional generator, the duality is justified by the fact that the functional generator of free Kalb-Ramond and Proca can be obtained from simple Gaussian integrals on $B_{\mu \nu}$ and $A_\mu$. We do not reproduce this argument here, essentially for two reasons. Since the dual Hamiltonians, free and interacting for our simple interaction, are related by canonical coordinate transformations it seems natural to use the parent action to relate dual Lagrangians also by coordinate transformations. Moreover, it is not straightforward to propose the correct parent action in the presence of interactions. Suppose we want to know what is the dual functional generator associated with the starting action \eqref{interacting-1}. In order to propose a parent action in the path integral, we would have to know beforehand that the action in the path integral is \eqref{PI_KR}. Then, we could propose the parent action  
    \begin{align}
         S_P  = \int d^4 x \Bigg[ - \frac{m^2}{2}A_\mu A^\mu + \frac{m^2}{4} B_{\mu \nu} B^{\mu \nu} - m B^{\mu \nu} \partial_\mu A_\nu + \frac{\lambda \phi^2 }{4}  B_{\mu \nu} B^{\mu \nu} 
         -\frac{\partial_\mu \phi \partial^\mu \phi}{2}\nonumber\\ - \frac{3 \, i }{2} \delta^4(0) \ln \left(1+ \lambda \phi^2 /m^2 \right) \Bigg],
    \end{align}
    in the path integral so that we can reproduce \eqref{PI_KR} and \eqref{PI_P} by functional integrals on $A_\mu$ and $B_{\mu \nu}$ respectively. This seems an ad-hoc procedure, the use of the parent action to relate dual Lagrangians through coordinate transformations seems more straightforward.
    
    \section{Discussion}
    
        It was shown that the canonical quantization of the free massive Kalb-Ramond and Curtright-Freund Lagrangian leads to a theory of massive spin-1 and spin-0 particles. These are the same theories obtained by canonical quantization of the Proca and Klein-Gordon Lagrangians. We stressed that the consequence of the free equivalence is the duality of Feynman rules. Given an interacting Hamiltonian $\mathcal{H}(x)$, in the interaction picture, derived from a Kalb-Ramond or Curtright-Freund Lagrangian, it is expected that the same $\mathcal{H}(x)$ can be derived starting with a Proca or Klein-Gordon Lagrangian, respectively. An interacting Kalb-Ramond Lagrangian was proposed and we checked that there is a dual Proca Lagrangian so that both define the same $\mathcal{H}(x)$. We derived the covariant Feynman rules in the canonical and path integral formalism. We emphasized in the path integral formalism that the Feynman rules derived from a dual model like Kalb-Ramond or Curtright-Freund are equivalent to the Feynman rules derived from a Proca or Klein-Gordon Lagrangian once we sum the effect of contact terms. Finally, we showed that for the simple interaction in this paper the KR-P duality goes beyond perturbation theory, the interacting Hamiltonians derived from Kalb-Ramond and Proca Lagrangians are related by canonical coordinate changes. We also introduced the parent action method to argue that the non-perturbative equivalence may be valid for more general interactions. Our non-standard use of the parent action is because the Hamiltonians of dual models are related by coordinate changes so it seems more natural to also relate their Lagrangians by coordinate changes. Moreover, it is not straightforward to propose a parent action in the presence of interactions.

        Although we have focused on Kalb-Ramond and Curtright-Freund models, we expect this analysis to be valid for dual massive models without gauge symmetry in general. As stressed in \ref{apen1} the source of dual massive models is a coordinate change: the Kalb-Ramond and Proca Lagrangians are Legendre transforms of the same Hamiltonian in different canonical coordinates. Naturally, such coordinate transformations should be possible for higher spins. For spin-1 there is already another choice of canonical coordinates that leads to a description in terms of a symmetric tensor \cite{Dalmazi:2011df}. The parent action method suggests that this interpretation is maintained in the presence of interactions: dual Lagrangians are derived from the same parent action by coordinate changes. It is a reasonable expectation that the Hamiltonians derived from these Lagrangians are also related by coordinate changes.  However, we do not know of any theorem stating that coordinate changes in the Lagrangian formalism always correspond to canonical coordinate changes in the Hamiltonian formalism. It would be useful to have such a theorem if we want to take dual massive Lagrangians as the starting point of interacting theories. The coordinate changes connecting Proca to Kalb-Ramond or Curtright-Freund to Klein-Gordon are not defined for zero mass, so in the massless limit the theories are not related, which is a general feature of dual massive models.

        Regarding the recent work \cite{Hell:2021wzm} there is an important point that should be made. In this work, self-interacting Lagrangians were compared and it was pointed out that the self-interacting theories of Kalb-Ramond and Proca are not equivalent. We would like to remark that the content of KR-P duality is not between any two Lagrangians of Kalb-Ramond and Proca and it is not clear whether the proposed self-interacting Lagrangians should be dual in the first place. The interaction terms are similar, but even for simple interactions we learned from \eqref{interacting-1} and \eqref{interacting-2} that the Kalb-Ramond and Proca Lagrangians that define the same content, non-perturbatively, can have radically different interactions. The dual Lagrangians are supposed to be derived from the same parent-action, which does not seem to be possible for the suggested self-interacting theories.

\section{Acknowledgements}
I am grateful to my advisor D. Dalmazi for motivating me to investigate the duality in the context of the path integral, for reading and comment the draft, and for useful discussions on this work. I would like to thank also to J. M. Hoff da Silva for his comments on the draft and R. R. Lino dos Santos for carefully reading the draft and precise suggestions. This work was supported by CAPES-Brazil.

\appendix
\section{Canonical formalism backward}

\subsection{Kalb-Ramond}\label{apen1}

Our starting point is the Hilbert space of massive spin-1 particles according to Wigner's classification \cite{Wigner:1939cj}, that is the one particle states $| \boldsymbol{p} \, \sigma \rangle$ furnish an irreducible representation of the Lorentz group, with $\sigma = -1,0,1$, and the normalization adopted here is
\begin{align}
    \langle \boldsymbol{p}' , \sigma' | \boldsymbol{p} , \sigma \rangle =  \delta_{\sigma' \, \sigma} \delta^3 (\boldsymbol{p}' - \boldsymbol{p}).
\end{align}
We define the boost $L^\mu\,_\nu(p)$ which takes the particle with momentum $k^\mu = (m, \boldsymbol{0})$ to momentum $p^\mu =(\sqrt{\boldsymbol{p}^2+m^2}, \boldsymbol{p})$:
\begin{align}
    | \boldsymbol{p}, \sigma \rangle &= \sqrt{\frac{k^0}{p^0}} U\left( L(p) \right) | \boldsymbol{0}, \sigma \rangle\\
    L^0\,_0 (p) = \gamma, \qquad L^0\,_i(p) = L^i\,_0(p) &= \hat{p}^i \sqrt{\gamma^2-1}, \qquad L^i \,_j (p) = \delta_{ij} + (\gamma-1) \hat{p}^i \hat{p}^j ,
\end{align}
where $\gamma \equiv p^0/m$ and $\hat{p}^i \equiv p^i / | \boldsymbol{p} |$. The effect of an general homogeneous Lorentz transformation is 
\begin{align}
 U \left( \Lambda \right) | \boldsymbol{p} , \sigma \rangle = \sqrt{ \frac{(\Lambda p )^0}{p^0}} \sum_{\sigma'} D_{\sigma' \sigma }\left( W(\Lambda, p) \right) | \boldsymbol{p}_\Lambda , \sigma' \rangle ,
\end{align}
where $D_{\sigma' \sigma }$ is the spin-1 matrix representation of the Wigner rotation:  $W(\Lambda,p) \equiv L^{-1}(\Lambda p ) \Lambda L(p)$. We define the creation and annihilation operators $a(\boldsymbol{p}, \sigma) \, , \, a^{\dagger}(\boldsymbol{p},\sigma)$ so that
\begin{subequations}
\begin{align}
    &\left[ a(\boldsymbol{p}, \sigma) \, , \, a^{\dagger}(\boldsymbol{p}',\sigma') \right] = \delta_{\sigma  \sigma'} \delta^3(\boldsymbol{p}- \boldsymbol{p}'), \qquad a(\boldsymbol{p}, \sigma) | 0 \rangle = 0,\\
    &U(\Lambda) a^{\dagger} (\boldsymbol{p}, \sigma) U^{-1} (\Lambda) = \sqrt{\frac{(\Lambda p)^0}{p^0}} \sum_{\sigma'} D_{\sigma' \sigma }\left( W( \Lambda, p) \right) a^{\dagger} (\boldsymbol{p}_\Lambda, \sigma'),
\end{align}
\end{subequations}
where $|0\rangle $ is the vacuum state. The action of creation operators in $|0 \rangle$ leads to several particle states. The Hamiltonian is just \eqref{hamilfs=1}.

In order to implement the canonical formalism the first task is to define canonical variables $q_i (\boldsymbol{x},t)$ and $\pi^i(\boldsymbol{x},t)$ that satisfy 
\begin{subequations}\label{canvariadef}
\begin{align}
    \left[ q_i (\boldsymbol{x},t) \, , \, \pi^j (\boldsymbol{y},t) \right] &= i \, \delta^j_i \, \delta^3 (\boldsymbol{x}- \boldsymbol{y} ), \\
    \left[ q_i (\boldsymbol{x},t) \, , \, q_j (\boldsymbol{y},t) \right] &= \left[ \pi^i (\boldsymbol{x},t) \, , \, \pi^j (\boldsymbol{y},t) \right] = 0,
\end{align}
\end{subequations}
and in such a way that \eqref{hamilfs=1} can be rewritten as a functional of these variables.

 Given a set of canonical variables that satisfy \eqref{canvariadef} for any functional $F[q(t),\pi(t)]$ \footnote{We keep the time argument in $q_i (\boldsymbol{x},t)$ and $\pi^i(\boldsymbol{x},t)$ to indicate that in the functional $F$the spatial variables are integrated out at fixed time.} we can define the quantum mechanical functional derivatives
\begin{align}
    \frac{\delta F[q(t),\pi(t)]}{\delta q_i (\boldsymbol{x},t )} \equiv i \left[ \pi^i (\boldsymbol{x},t) \, , \, F[q(t),\pi(t)] \right] ,\\
    \frac{\delta F[q(t),\pi(t)]}{\delta \pi^i (\boldsymbol{x},t)} \equiv i \left[ F[q(t),\pi(t)] \, , \, q_i (\boldsymbol{x},t) \right].
\end{align}
This definition is motivated by the fact that if $F[q(t),\pi(t)]$ is written with all the q's to the left of all the p's taking the functional derivative is equivalent to taking the commutator. For instance, if 
\begin{align}
    F[q(t),\pi(t)] =  \int d^3 x \, q_1 (\boldsymbol{x},t) \,  q_3(\boldsymbol{x},t) \, \pi^2(\boldsymbol{x},t) \pi^3(\boldsymbol{x},t) , 
\end{align}
then 
\begin{align}
    \frac{\delta F[q(t),\pi(t)]}{\delta q_1 (\boldsymbol{y},t )} &= \int \, d^3 x \, \delta^3 (\boldsymbol{x} - \boldsymbol{y} ) q_3(\boldsymbol{x},t) \, \pi^2(\boldsymbol{x},t) p^3(\boldsymbol{x},t) \\
    &= \int d^3 x \, i \left[ \pi^1(\boldsymbol{y},t) \, , \, q_1(\boldsymbol{x},t) \right] q_3(\boldsymbol{x},t) \, \pi^2(\boldsymbol{x},t) \pi^3(\boldsymbol{x},t).
\end{align}
The implementation of the canonical formalism is  complete when we check that our variables satisfy the canonical equations of motion
\begin{subequations}\label{hamformeqm}
\begin{align}
    \dot{q}_i (\boldsymbol{x},t) &= \frac{\delta H[q(t),\pi(t)]}{\delta \pi^i(\boldsymbol{x},t)} = i \left[ H , q_i(\boldsymbol{x},t) \right],\\
    \dot{\pi}_i (\boldsymbol{x},t) &= - \frac{\delta H[q(t),\pi(t)]}{\delta q^i(\boldsymbol{x},t)} = i \left[ H , q_i(\boldsymbol{x},t) \right].
\end{align}
\end{subequations}

One possibility for implement the canonical formalism is in terms of the Proca variables, explored in \cite{weinberg_1995}:
\begin{align}
   & q^P_i (\boldsymbol{x},t) = (2 \pi)^{-3/2} \sum_{\sigma}   \int \frac{d^3 p}{(2 p^0)^{1/2}}
  \Bigg[ e_i (\boldsymbol{p}, \sigma ) a( \boldsymbol{p}, \sigma) e^{i p \cdot x}  + e^*_i  (\boldsymbol{p}, \sigma)  a^{\dagger} (\boldsymbol{p}, \sigma) e^{-i p \cdot x} \Bigg], \\
  &\pi_P^i (\boldsymbol{x},t)  = i \, (2 \pi)^{-3/2} \sum_{\sigma}   \int \frac{d^3 p}{(2 p^0)^{1/2}}
  \Bigg[ p_{[0} e_{i]} (\boldsymbol{p}, \sigma ) a( \boldsymbol{p}, \sigma) e^{i p \cdot x} 
  - p_{[0} e^*_{i]}  (\boldsymbol{p}, \sigma)  a^{\dagger} (\boldsymbol{p}, \sigma) e^{-i p \cdot x} \Bigg]  ,
\end{align}
in which case \eqref{hamilfs=1} is 
\begin{align}\label{hamiproca}
    H_P = \int \frac{d^3 x}{2} \left[ \boldsymbol{\pi}_P^2 + \frac{(\boldsymbol{\nabla} \cdot \boldsymbol{\pi}_P)^2}{m^2} + (\boldsymbol{\nabla} \times \boldsymbol{q}_P)^2 + m^2 \boldsymbol{q}_P^2 \right],
\end{align}
up to the constant term $\delta^3(\boldsymbol{0})$. However, we could also have choose
\begin{align}\label{canovarKR}
    q_i (\boldsymbol{x},t) = m^{-1} \pi_P^i (\boldsymbol{x},t), \qquad
  \pi^i (\boldsymbol{x},t)  = -m q^P_i (\boldsymbol{x},t) ,
\end{align}
in this case \eqref{hamilfs=1} reads
\begin{align}\label{apenKRH}
H  =  \int \frac{d^3 x}{2} \left[ \boldsymbol{\pi}^2 + (\boldsymbol{\nabla} \cdot \boldsymbol{q})^2 +  \frac{(\boldsymbol{\nabla} \times \boldsymbol{\pi})^2}{m^2} + m^2 \boldsymbol{q}^2 \right],
\end{align}
again up to a constant term. The set of variables \eqref{canovarKR} and the Hamiltonian \eqref{apenKRH} satisfy \eqref{canvariadef} and \eqref{hamformeqm}, and therefore are a valid choice of canonical coordinates. From the Lorentz transformation properties of $e^{\mu}(\boldsymbol{p}, \sigma)$, we note that the quantity 
\begin{align}
  B^{[\mu \nu]} \propto  \frac{i}{m} (2 \pi)^{-3/2} \sum_{\sigma} \int d^3 p (2 p^0)^{-1/2} \left[ p^{[\mu} e^{\nu]}(\boldsymbol{p}, \sigma ) a( \boldsymbol{p}, \sigma) e^{i p \cdot x} 
      - p^{[ \mu}e^{* \nu]} (\boldsymbol{p}, \sigma)  a^{\dagger} (\boldsymbol{p}, \sigma) e^{-i p \cdot x} \right],
\end{align}
is a representation of the Lorentz group. Motivated by that we are led to define $B_{[0i]} \equiv  q_i$ and introduce the auxiliary variable \begin{align}
    B_{[ij]} \equiv -\frac{1}{m^2} \left( \partial^i \pi^j - \partial^j \pi^i \right).
\end{align}
Similar considerations for the Proca variables would lead to the definition $A^i \equiv q^P_i $, and the introduction of the auxiliary variable 
\begin{align}
    A^0 =  \frac{\boldsymbol{\nabla} \cdot \boldsymbol{\pi}_P}{m^2}.
\end{align}
 The Hamiltonians \eqref{hamiproca} and \eqref{apenKRH} in terms of the auxliar variables are 
\begin{subequations}
\begin{align}\label{APEN-HP}
    H_P = \int \frac{d^3 x}{2} \left[ \boldsymbol{\pi}_P^2 + m^2 A_0^2 + (\boldsymbol{\nabla} \times \boldsymbol{A})^2 + m^2 \boldsymbol{A}^2 \right], \\ \label{APEN-KR}
    H =  \int \frac{d^3 x}{2} \left[ \boldsymbol{\pi}^2 + (\partial^l B_{[0l]})^2 +  \frac{m^2}{2} B^2_{[ij]} + m^2 B^2_{[0l]} \right]
\end{align}
\end{subequations}
The Lagrangian is defined as the Legendre transformation 
\begin{align}
    L[q(t), \dot{q}(t) ] \equiv \int d^3 x \, \pi^j (\boldsymbol{x},t) q_j (\boldsymbol{x},t)  -  H[ q(t), \pi(t) ],
\end{align}
with the understanding that $\pi^j(\boldsymbol{x},t)$ is to be expressed as a function of $q_j(\boldsymbol{x},t)$ and the auxiliary variables. The Legendre transformations of \eqref{APEN-HP} and \eqref{APEN-KR} are, respectively 
\begin{subequations}
\begin{align}
    &L_P = \int d^3 x \left[- \frac{1}{4}  F_{\mu \nu} F^{\mu \nu} - \frac{m^2}{2} A_\mu A^\mu  \right]\\
    &L_{KR} = \int d^3 x \left[\frac{\left(\partial^\mu B_{[\mu \nu]} \right)^2 }{2} + \frac{m^2}{4} B_{[\mu \nu]} B^{[\mu \nu]} \right]
\end{align}
\end{subequations}
 For Proca and Kalb-Ramond, we introduced the auxiliary variables by the search of a tensor so that the canonical variables $q$ are components of this tensor. This is motivated by our final goal, which is a manifestly Lorentz invariant function $L\left[ q(t), \dot{q}(t) \right]$.

\subsection{Curtright-Freund}\label{apens=0}

We can use all the previous definitions for spin-0. The only qualification is that there is no $\sigma$ index. In complete analogy with the previous section we can implement the canonical formalism with the Klein-Gordon variables
\begin{align}
   & q^{KG} (\boldsymbol{x},t) \equiv (2 \pi)^{-3/2}  \int
    \frac{d^3 p }{(2 p^0)^{1/2}} \left(  a( \boldsymbol{p}) e^{i p \cdot x} 
      +   a^{\dagger} (\boldsymbol{p}) e^{-i p \cdot x} \right),\\
      & \pi^{KG} (\boldsymbol{x},t) \equiv i \,  (2 \pi)^{-3/2}  \int
    \frac{d^3 p }{(2 p^0)^{1/2}} p_0 \left(  a( \boldsymbol{p}) e^{i p \cdot x} 
      -   a^{\dagger} (\boldsymbol{p}) e^{-i p \cdot x} \right),
\end{align}
so that 
\begin{subequations}
\begin{align}
   & \left[ q^{KG}( \boldsymbol{x}, t ) \, , \, \pi^{KG} ( \boldsymbol{y} ,t ) \right]  =  i \delta^3 ( \boldsymbol{x} - \boldsymbol{y} ),\\
   & \left[ q^{KG} (\boldsymbol{x}, t) \, , \, q^{KG} (\boldsymbol{y}, t) \right]  =  \left[ \pi^{KG} ( \boldsymbol{x} ,t ) \, , \, \pi^{KG}( \boldsymbol{y} ,t ) \right] = 0,
\end{align}
\end{subequations}
and the free Hamiltonian is 
\begin{align}\label{hamilKG}
    H = \int \frac{d^3x}{2} \left[ m^2 q_{KG}^2 + \pi_{KG}^2 + (\boldsymbol{\nabla} q_{KG})^2 \right].
\end{align}
On the other hand, we could also have choose the canonical variables 
\begin{align}
q(\boldsymbol{x},t) \equiv m^{-1} \pi^{KG}(\boldsymbol{x},t) , \qquad \pi(\boldsymbol{x},t) \equiv -m  q^{KG}(\boldsymbol{x},t),
\end{align}
in which case 
\begin{subequations}
\begin{align}
   & \left[ q( \boldsymbol{x}, t ) \, , \, \pi ( \boldsymbol{y} ,t ) \right]  =  i \delta^3 ( \boldsymbol{x} - \boldsymbol{y} ),\\
   & \left[ q (\boldsymbol{x}, t) \, , \, q (\boldsymbol{y}, t) \right]  =  \left[ \pi( \boldsymbol{x} ,t ) \, , \, \pi( \boldsymbol{y} ,t ) \right] = 0,\\\label{hamilorigi}
     & H = \int \frac{d^3x}{2} \left[ \pi^2 + m^2 q^2 + \frac{(\boldsymbol{\nabla} \pi)^2}{m^2} \right].
\end{align}
\end{subequations}
We note that the canonical variable $q_{KG}(\boldsymbol{x},t)$ is already a representation of the Lorentz group, there is no need to define auxiliary variables in performing the Legendre transformation of the Klein-Gordon Hamiltonian. That is not the case for $q(\boldsymbol{x},t)$. Noticing that
\begin{align}
    B_{\mu} (\boldsymbol{x},t) \propto    (2 \pi)^{-3/2}  \int
    \frac{d^3 p }{(2 p^0)^{1/2}} p_\mu \left(  a( \boldsymbol{p}) e^{i p \cdot x} 
      -   a^{\dagger} (\boldsymbol{p}) e^{-i p \cdot x} \right),
\end{align}
 we are led to define $B_0 \equiv q$ and introduce the auxiliary variable
\begin{align}
    B_i (\boldsymbol{x},t) = -  \frac{\partial_i \pi(\boldsymbol{x},t)}{m^2} .
\end{align}
Now \eqref{hamilorigi} reads 
\begin{align}\label{hamilorigi2}
    H = \int \frac{d^3x}{2} \left[ \pi^2 + m^2 q^2 + m^2 \boldsymbol{B}^2 \right].
\end{align}
The Legendre transformation of \eqref{hamilKG} and \eqref{hamilorigi2} are respectively
\begin{align}
    &L = \int \frac{d^3 x }{2} \left[ -(\partial_\mu q_{KG})^2 - m^2 q_{KG}^2 \right], \\
    &L = \int \frac{d^3 x }{2} \left[ (\partial_\mu B^\mu)^2 + m^2 B_\mu^2 \right].
\end{align}

\section{Path integral}\label{PI}

The Hamiltonian path integral associated to  \eqref{fullHInKR} and \eqref{reducedproca} is, respectively 
\begin{align}\label{hamPIKR}
     &\langle Vac,in |  Vac,out \rangle =
      \int \mathcal{D} \phi \mathcal{D} \pi \mathcal{D} B_{0 i} \mathcal{D} \Pi^{0i} \, e^ {   i \int d \tau \left(\int d^3 x \, \dot{\phi} \pi + \dot{B}_{0i} \Pi^{0i} - H^{R}_{KR} \right)  }, \\ \label{hamPIP} 
    &\langle Vac,in |  Vac,out \rangle =  \int \mathcal{D} \phi \mathcal{D} \pi  \mathcal{D} A_{i} \mathcal{D} \bar{\Pi}^i \, e^{   i \int d \tau \left( \int d^3 x \, \dot{\phi} \pi + \dot{A}_i \bar{\Pi}^i - H^R_P \right) },
\end{align}
where $H^{R}_{KR}$ and $H^R_P$ are defined in \eqref{fullHInKR} and \eqref{reducedproca}. We note that 
\begin{align}
    &I_1 \equiv \int \mathcal{D} B_{ij} \, e^{i \int d^4 x \frac{m^2}{4} \left( 1 + \lambda \phi^2/m^2 \right) \left[  B_{ij} + \frac{1}{m^2} \left( 1 + \lambda \phi^2/m^2 \right)^{-1} \left(\partial^i \Pi^{0j} - \partial^j \Pi^{0i} \right) \right]^2   } = \mathcal{N} \, e^{- \frac{3 \delta^4(0)}{2} \int d^4 x \ln\left(1 + \lambda \phi^2/m^2 \right)     }, \nonumber\\
    &I_2 \equiv \int \mathcal{D} A_0 \, e^{i \int d^4 x \frac{m^2}{2} \left( A_0 + \frac{\boldsymbol{\nabla} \cdot \bar{\boldsymbol{\Pi} } }{m^2 } \right)^2 } = \mathcal{N}' , \nonumber
\end{align}
where $\mathcal{N}, \mathcal{N}'$ are field independent constants. We reintroduce auxiliary variables in \eqref{hamPIKR} and \eqref{hamPIP} using $I_1 \, (I_1)^{-1} = I_2 \, (I_2)^{-1}  = 1$ :
\begin{align}
    & \langle Vac,in |  Vac,out \rangle = \mathcal{N}^{-1}
      \int \mathcal{D} \phi \mathcal{D} \pi \mathcal{D} B_{\mu \nu} \mathcal{D} \Pi^{0i} \, e^ {   i \int d \tau \left[\int d^3 x \left( \dot{\phi} \pi + \dot{B}_{0i} \Pi^{0i} - \frac{3\, i \delta^4(0)}{2}  \ln \left(1+ \lambda \phi^2 /m^2 \right) \right) - H_{KR} \right]  }   ,\nonumber \\ 
   & \langle Vac,in |  Vac,out \rangle = (\mathcal{N}')^{-1} \int \mathcal{D} \phi \mathcal{D} \pi  \mathcal{D} A_{\mu} \mathcal{D} \bar{\Pi}^i \, e^{   i \int d \tau \left( \int d^3 x \, \dot{\phi} \pi + \dot{A}_i \bar{\Pi}^i - H_P \right) }, \nonumber
\end{align}
where 
\begin{align}
   & H_{KR} = \int d^3 x  \bigg[ \frac{1}{2} \pi^2 + \frac{1}{2} \Pi^{0 i } \Pi^{0 i } + \Pi^{0 i } \partial^j B_{j i } + \frac{1}{2} (\boldsymbol{\nabla} \phi )^2  + \frac{1}{2} (\partial^i B_{0i})^2 -  \frac{1}{4} (m^2 + \lambda \phi^2) B_{\mu \nu} B^{\mu \nu} 
   \bigg], \nonumber \\
   & H_{P} = \int d^3 x \Bigg[ \left( 1 + \frac{\lambda \phi^2 }{m^2 } \right) \frac{\bar{\boldsymbol{\Pi}}^2}{2} + \frac{\pi^2}{2} - A_0 \boldsymbol{\nabla} \cdot \bar{\boldsymbol{\Pi}} + \left( 1 + \frac{\lambda \phi^2 }{m^2 } \right)^{-1} \frac{(\boldsymbol{\nabla} \times \boldsymbol{A} )^2 }{2} + \frac{(\nabla \phi)^2}{2} 
    + \frac{m^2 }{2} A_\mu A^\mu \Bigg]. \nonumber
\end{align}
The Gaussian integral over the momenta just recovers the Lagrangian path integral 
\begin{align}
     &\langle Vac,in |  Vac,out \rangle =
     \mathcal{N}^{-1} \mathcal{K} \int \mathcal{D} \phi \mathcal{D} B_{[\mu \nu]} \, e^{   i \int d^4 x \left[ \mathcal{L}^0_{KR}
    + \frac{\lambda}{4} \phi^2 B_{\mu \nu} B^{\mu \nu}  -\frac{3 \, i }{2} \delta^4(0) \ln \left(1+ \lambda \phi^2 /m^2 \right)\right]  }, \\ 
    &\langle Vac,in |  Vac,out \rangle = (\mathcal{N}')^{-1} \mathcal{K}' \int \mathcal{D} \phi \mathcal{D} A_{\mu} \, e^{   i \int d^4 x \left[ \mathcal{L}^0_{P}
    + \frac{\lambda \phi^2}{4 m^2}  \frac{F_{\mu \nu} F^{\mu \nu}}{\left(1+ \lambda \phi^2/m^2\right)}  + \frac{3 \, i }{2} \delta^4(0) \ln \left(1+ \lambda \phi^2 /m^2 \right)\right]    }
\end{align}
where $\mathcal{K}, \mathcal{K}'$ are also field independent constants and 
\begin{align}
    \mathcal{L}^0_{KR} &\equiv - \frac{1}{2} \partial_\mu  \phi \partial^\mu \phi + \frac{1}{2} \partial_\alpha B^{\alpha \mu} \partial^{\nu} B_{\nu \mu} + \frac{m^2}{4} B_{\mu \nu} B^{\mu \nu}, \\
    \mathcal{L}^0_P &\equiv - \frac{1}{2} \partial_\mu \phi \partial^\mu\phi - \frac{1}{4} F_{\mu \nu} F^{\mu \nu} - \frac{m^2}{2} A_\mu A^\mu .
\end{align}

\end{document}